# Scalable Resource Provisioning for Multi-user Communications in Next Generation Networks

A. Neto
Institute of Telecommunications
Aveiro, Portugal
augusto@av.it.pt

E. Cerqueira, M. Curado and E. Monteiro
University of Coimbra
Coimbra, Portugal
{ecoelho; marilia; edmundo}@dei.uc.pt

P. Mendes
INESC
Porto, Portugal
pmendes@inescporto.pt

*Abstract*—The great demand for real-time multimedia sessions encompassing groups of users (multi-user), associated with the limitations of the current Internet in providing quality assurance, has raised challenges for defining the best mechanisms to deploy the *Next Generation of Networks* (NGN). There is a consensus that an efficient and scalable provisioning of network resources is crucial for the success of the NGN, mainly in what concerns access networks. Previous solutions for the control of multi-user sessions rely mostly on uncoordinated actions to allocate per-flow bandwidth and multicast trees. This paper introduces a *Multi-user Aggregated Resource Allocation* mechanism (MARA) that coordinates the control of class-based bandwidth and multicast resources in a scalable manner. In comparison with previous work, MARA significantly reduces signaling, state and processing overhead. The performance benefits of MARA are analyzed though simulations, which successfully demonstrated the significant optimization in the network performance.

## I. INTRODUCTION

Changes in competition patterns are forcing *Internets Service Providers* (ISPs) to support more competitive applications. Some of those applications rely on the simultaneous delivery of content to groups of users (multi-user sessions), such as (mobile) IPTV, video streaming and video conferencing. In this context, the use of IP multicast allows the deployment of multi-user communications with good usage of resources in constrained (access) IP networks, as done today by DSL access providers. However, the per-flow control of IP multicast trees associated with the lack of coordination between the control of multicast and bandwidth resources may limit the success of large scale deployment of multi-user IP communications in the *Next Generation of Networks* (NGN).

Trying to overcome some of the described limitations, the authors proposed in a previous work a mechanism, called *MultI-Service Resource Allocation* (MIRA) [1], supporting the coordinated control of Source-Specific Multicast (SSM) trees and bandwidth of DiffServ classes. With MIRA, the complexity of resource allocation is pushed to network edges, while interior routers remain simple. MIRA coordinates resource allocation within a network via a per-flow edge-to-edge ingress-driven signaling approach implemented by the *MIRA Protocol* (MIRA-P). MIRA-P uses only two messages for resource requests (indicated by message-specific flags) and operation feedback, called *RESERVE* and *RESPONSE* respectively. The performance evaluation presented in [1] shows that MIRA distinguishes itself from existing solutions by providing a low complexity approach to minimize signaling and state overhead. However, the main limitation of MIRA resides in the per-flow signaling to control the bandwidth of DiffServ classes and IP multicast trees. Whereas the control of bandwidth introduces excessive signaling load and processing overhead in large-scale networks, the control of multicast trees suffers from state overhead and has no access control and QoS support. Our performance findings in the analysis of MIRA motivated further investigation of methods to avoid per-flow operations, as much as possible.

This paper extends our previous work by proposing the *Multi-user Aggregated Resource Allocation* (MARA), aiming to minimize the signaling, state and processing overhead of MIRA per-flow operation. MARA dynamically controls the over-provisioning of DiffServ classes bandwidth and multicast trees to increase the scalability of IP multi-user sessions on NGNs. The combination of surplus resources (bandwidth and multicast trees) with admission control allows MARA to establish multiple multi-user sessions without per-flow signaling. Multicast aggregation is used to optimize multicast state storage. To analyze the expected benefits of MARA over MIRA, a performance evaluation is carried out by simulations. The main goal is to analyze bandwidth and multicast state, as well as signaling overhead.

The remainder of this paper is organized as follows. Related work is presented in section 2. Section 3 provides a detailed description of MARA, whose performance is evaluated in Section 4. Finally, Section 5 concludes this paper with a summary of our findings.

## II. RELATED WORK

Resource allocation based on over-reservations has been addressed in the literature on a set of proposals. The *Border Gateway Reservation Protocol* (BGRP) [2] and the *Shared-segment Inter-domain Control Aggregation* protocol (SICAP) [3], use a similar two-pass receiver-driven signaling schema to

This work was done at the Laboratory of Communications and Telematics of the Faculty of Science and Technology of the University of Coimbra. It is supported by DoCoMo Euro-labs, by the Portuguese Ministry of Science, Technology and High Education, and by European Union FEDER - POSI (projects Q3M and SAPRA). At the time, Augusto Neto was working in University of Coimbra, and Paulo Mendes in NTT DoCoMo Euro-Labs.

aggregate unicast traffic into sink- and shared trees respectively. That is, neither of the two approaches aims to handle multicast trees. Nevertheless, SICAP shows that a dynamic management of surplus reservations brings more benefits than the fixed quantification factor used by BGRP.

The *Dynamic Aggregation of Reservations for Internet Services* (DARIS) [4] is a centralized approach in which over-reservations are deployed based on knowledge of the internal topology, resource capacities and current selected routes. Thus, scalability is the main issue in DARIS, since centralization endangers system performance and requires excessive control. The *Simple Inter-Domain QoS Signaling Protocol* (SIDSP) [5] establishes flows based on over-reservations, hard-state resource maintenance and source routing. The authors limit to say that over-reservations can be setup based on previously network events. Moreover, an additional mechanism is needed to control state coherence, addressing the problems associated with the hard-state. SIDSP also suffers from scalability problems due to the two-pass signaling schema, as well as due to the addition of IP addresses in the header of all data packets for source-routing. In the latter, the increasing packet size (dependent upon the number of hops in the communication path) introduces bandwidth and processing overhead.

In what concerns the over-provisioning of multicast trees, there is a lack of efficient solutions, since most of them aim to protect the system against topology changes (e.g., link breaks), instead of reducing signaling overhead to setup and maintain new sessions. Some solutions [6], [7] provide backup multicast trees and controls switching between main and backup trees dynamically. The signaling overhead is not optimized, since the system must be signaled whenever a new multicast tree is required. In spite of speeding up the restoration of sessions affected by re-routing, these solutions can introduce serious scalability problems in large-scale systems since to each multicast tree, a backup tree is required.

The related work analysis shows that none of the proposals efficiently supports the coordinated over-provisioning of network bandwidth and multicast trees. Most of the approaches supporting the over-reservation of bandwidth have performance limitations (two-pass signaling approach) and are not efficient to multi-user sessions (no multicast support). Additionally, the multicast-aware proposals create surplus multicast trees in advance for reliability proposes and not to optimize signaling overhead. Thus, MARA is proposed to address the above challengers.

III. MULTI-USER AGGREGATED RESOURCE ALLOCATION

MARA deploys dynamic over-provisioning of network resources (bandwidth and multicast) without per-flow signaling. Quality of service is assured based on DiffServ per-class over-reservations, being resources dynamically re-adjusted to map bandwidth utilization. Multicast trees are over-provisioned in advance (at the system bootstrap) and dynamically connected based on multicast aggregation. Thus, MARA expects to minimize the signaling and processing overhead of MIRA and other per-flow signaling approaches.

MARA re-uses some of the components developed for MIRA, namely the signaling protocol (MIRA-P) and the *Resource Control Function* (RCF). MIRA-P provides control information so that RCF can control the state of network elements to setup network resources. In relation with MIRA, the common header of MIRA-P messages was extended with two new message-specific flags to invoke the initialization of network resources (flag *Initialization* (I)) and dynamically adjust over-reservations to current demands (flag *Over-reservation* (O)).

The *Advanced QoS Resource Allocator* (ASAC) and the *Advanced Aggregation Tree Allocator* (AGTree) are conceived to control the over-provision of bandwidth and multicast trees, respectively. Fig. 1 shows MARA components implemented in edge nodes of a network. Interior nodes only implement MIRA-P and RCF.

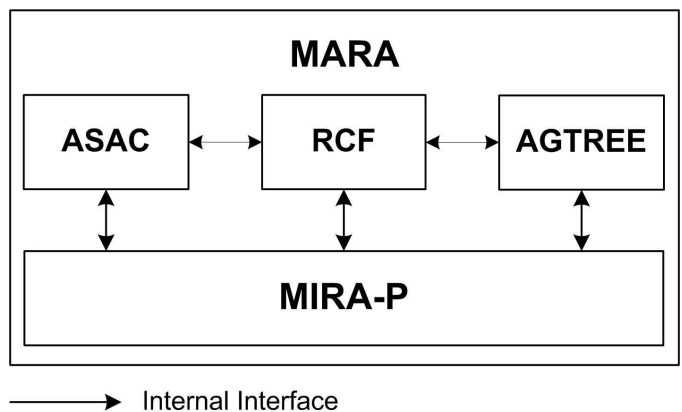

*Fig. 1.* Architecture of MARA

ASAC over-provisions bandwidth by assigning per-class over-reservations at the system bootstrap and controlling their adjustment dynamically without per-flow signaling. In order to avoid class starvations, each *Class of Service* (CoS) is assigned a *Committed* and a *Maximum Reservation Threshold* ($CRth$ and $MRth$, respectively), and a global initialization factor as a fraction of the local link capacity (e.g., ½, ¼ etc.). Whenever realized that a CoS is congested, ASAC tries to increase its over-reservation based on network utilization ratios and commitments. Moreover, the size of the $MRth$ of the CoSs is also controlled, so that a flow can be admitted when detecting that the $MRth$ of its required CoS is congested (not allowed in MIRA).

AGTree over-provisions multicast trees by assigning surplus trees at the system bootstrap. AGTree uses encapsulation to merge multicast flows into aggregation trees (instead of source-routing proposals) without any signaling exchange inside the network. Thus, data packets are aggregated at ingress routers and de-aggregated at egress routers accordingly. Moreover, AGTree controls multicast connectivity scope by dynamically switching sessions from one multicast aggregation tree to another as a consequence of network dynamics (e.g., new *join/leave* requests or re-routing due to link failures).

*A. MARA Functionalities*

This section aims to introduce the functionalities supported by MARA to deploy the over-provisioning of network resources. It is assumed that all MIRA agents have information about whether the local node is an ingress, core or egress router.

*1) System Initialization*

Ingress routers use a flooding mechanism to request the initialization of bandwidth, by multiplying the index factor by the *MRth* of each CoS, and to collect information about the available shortest paths until all available edges nodes. The information about the edge-to-edge paths is used to setup the aggregated multicast trees.

The system initialization may be described as follows: ingress routers setup the over-reservation of each CoS on all local network interfaces, based on the maximum reservation thresholds and the initialization factor. Afterwards, a *RESERVE(I)* (notation used to a *RESERVE* message with flag *I*) is composed. In addition to the initialization factor and per-class *MRth*, the *RESERVE(I)* must be prepared to collect information about the composition of the resultant distribution path. Therefore, a copy of the *RESERVE(I)* is sent in all interior interfaces (except the inter-domain link), where each one has the IP address of the associated local outgoing interface included in the *Reserved Path* (RSVPATH) object. As a consequence of the flooding schema, a node can receive several copies of the same message. To avoid redundant operations, each node must do some verification before setting up resources. For instance: i) ensuring that per-class over-reservation is initialized only once when the CoS has no reservation state; ii) avoiding infinite signaling loops by dropping messages already carrying the IP address of a local network interface in the *RSVPATH* object. As a result of the latter, long trees are avoided by dropping *RESERVE(I)* messages the return to ingress routers.

After initiating bandwidth resources and updating the *RSVPATH* object, each core node sends a copy of the *RESERVE(I)* to all local network interfaces, except the one in which it was received. In the case of an edge router a *RESPONSE(OK)* message is composed with the information derived from the associated *RESERVE(I)*, after the local bandwidth initialization is concluded. The *RESPONSE(OK)* is sent to the IP address found in the first entry of the *RSVPATH* object, allowing the ingress router to store information about the edge-to-edge path and about the used bandwidth and thresholds of the CoS of the bottleneck link of all paths. Fig. 2 illustrates the operations of system initialization.

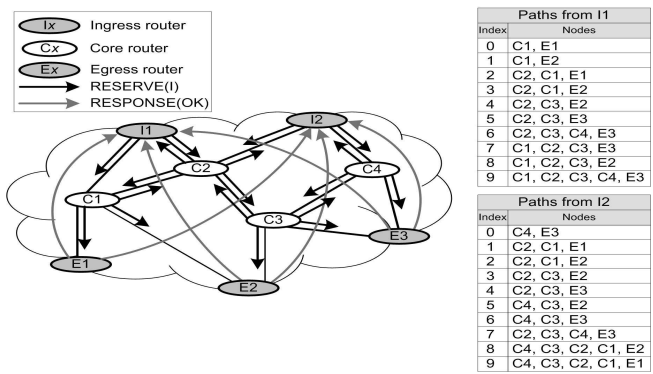

*Fig. 2.* Signaling events for the initialization of bandwidth resources

After successfully accomplished the initialization of CoS bandwidth, MARA setups the multicast trees along each collected communication path. As a first step, a SSM multicast channel is assigned to each communication path composed by a pair of ingress and egress routers (called un-branched). The channel is composed by the IP address of the ingress router and a multicast group IP. The multicast address is locally created by the ingress router by using dynamic multicast address allocation solutions [8]. After that, a *RESERVE(M)* (*RESERVE* message with a *Multicast* (M) flag) is sent downstream, being forwarded based on the *RSVPATH*. Each visited node configures the *Multicast Routing Information Base* (MRIB) with information derived from the *RSVPATH* object and propagates the message downstream. At the egress router, PIM-SSM is triggered to build the multicast tree, and a *RESPONSE(OK)* message is sent to the ingress router, confirming the successful operation.

After the creation of the un-branched trees, MARA uses a combinatorial algorithm to create a set of branched trees, simultaneously supplying multiple egress routers. At first all possible combinations between the available un-branched trees are generated in *n-1* interaction, where *n* is the number of un-branched trees. In order to optimize the large number of resultant branched trees, filters are used to retain the best ones. One selection criterion is based on the fact that a distribution tree demands more network resources (e.g., bandwidth) as branching points get close to the root. Based on this criterion, multicast trees with branching point in the ingress routers are discarded. Furthermore, combinations with multiple paths converging to the same node are also discarded. Optionally, MARA can be configured with a maximum number of hops that can be comprised in a multicast tree (this maximum number of hops can derive from network historical or measurement tools). Fig. 3 shows the branched and un-branched multicast trees generated by MARA in an illustrative scenario. After generated all multicast aggregation trees, the ingress routers signal the identified egress routers so that PIM-SSM can build the created multicast trees.

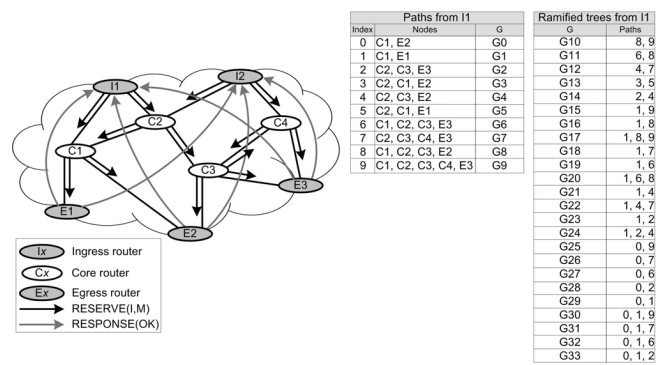

*Fig. 3.* List of all multicast aggregation trees created by MARA

The combinatorial algorithm can pose performance overhead to MARA, where the processing cost is overloaded with the increasing number of combinations (which increases with the number of links/nodes in the network). Other combinatorial algorithms can be used. The specification of an efficient combinatorial solution is out of the scope of this paper.

*2) Dynamic Resource Allocation*

The combination of over-provisioning and admission control allows MARA to dynamically allocate network resources without per-flow signaling. Whenever required the establishment of a session, MARA must be triggered at an ingress router with a session request (*Ri*) carrying the *QoS Specifications* (QSPEC) and the egress router IP. The QSPEC is composed at least by the CoS and the required bandwidth (*Brq(i)*). Based on the session definition, namely its egress router(s) and QoS requirements (CoS and available bandwidth), the ingress router selects one (the first matching) of the previously created multicast trees. After which, the session flow is encapsulated in identified multicast aggregation tree. At the egress router(s) associated to the multicast aggregation tree, the original IP header information is restored (de-aggregation).

If succeeding the information that the ingress router has about the current usage of the required CoS *i* in the bottleneck link of the required multicast tree, the session is admitted without further processing or signaling. On the other hand, if the required CoS *i* experiences unavailable bandwidth, ASAC attempts to re-adjust its current over-reservation. The amount of bandwidth to re-adjust the over-reservation of CoS *i* (*Bov(i)*) is given by (1). The equation (1) is based on the utilization ratio (*Bu(i)*) of the CoS and the amount of bandwidth which will be available after the session setup. The information used in (1) is related to the bottleneck link of the selected multicast aggregation tree. After the computation of a positive *Bov(i)* (in case of failure the equation returns a negative valour), MARA updates the current reservation of CoS *i*, *Brv(i)*, where *Brv(i)* ← *Bov(i)* + *Brq(i))*, and sends a *RESERVE(O)* in the selected communication path.

$$Bov(i) = \overbrace{\frac{Bu(i)}{MRth(i)}}^{\text{Utilization ratio}} * \overbrace{(MRth(i) - Bu(i) - Brq(i))}^{\text{Bandwidth available}} \quad (1)$$

All routers along the communication path update the local *Brv(i)* with the *Brq(i)* derived from the QSPEC transported in the *RESERVE(O)* message. Upon receiving the *RESPONSE(OK)*, the ingress router re-processes *Ri*, which is now supposed to succeed without signaling.

If the amount of resources requested by *Ri* exceeds the current *MRth(i)*, (1) does not succeed (*Bov(i)* < 0) and so the adjustment of the CoS over-reservation fails. In this case, MARA invokes the re-adjustment of the CoSs to admit *Ri* even under congestion. For this propose, (2) provides a bandwidth index (*B_Idx(j)*) of any other CoS *j* except the congested one (*i*). The bandwidth index of a CoS is the ratio between the amount of bandwidth that is currently reserved and used by CoS *j*.

$$\forall j \in c, B\_Idx(j) = \left(\frac{Brv(j) - Bu(j)}{Brv(j)}\right) \quad (2)$$

Additionally, the threshold index (*Th_Idx(j)*) of any CoS *j*, except the congested one (*i*), is provided by (3) as being the ratio between the maximum reservation threshold and a bandwidth reference (*Bref(j)*). The bandwidth reference is either the bandwidth currently reserved or the *CRth* of CoS *j* (if the *Bu(i)* is lesser than the *CRth(i)*). Thus, MARA ensures at least the QoS commitments of the classes.

$$\forall j \in c, Th\_Idx(j) = \left(\frac{MRth(j) - Bref(j)}{MRth(j)}\right) \quad (3)$$

The result of (2) and (3) are used to compute, by means of (4), the amount of bandwidth by which the *MRth* of each CoS *j* (*Brl_MRth(j)*) will be reduced. Equation (4) computes *Brl_MRth(j)* based on the average between the sum of bandwidth and threshold indexes of CoS *j*, multiplied by its amount of bandwidth currently available.

$$\forall j \in c, Brl\_MRth(j) = \left(\frac{B\_Idx(j) + Th\_Idx(j)}{2}\right) * (MRth(j) - Bref(j)) \quad (4)$$

After the computation of (4), MARA adds *MRth(i)* with the sum of the *Brl_MRth(j)* of the selected CoS (*j*), and decreases the *MRth(j)* of each CoS (*j*) by the computed value. After successfully accomplished the CoS re-adjustment, *Ri* is re-processed again. If the re-adjustment of the CoSs fails *Ri* is rejected. As in MIRA, MARA allows explicit releasing of resources, which is done by signaling the network with a *RESERVE(T)* (*RESERVE* message with *Tear* (T) flag)

*3) Session Connectivity Control*

MARA ensures the continuity of on-going sessions during their entire lifetime by automatically switching them between available multicast aggregation trees. This operation is deployed at the ingress router associated with the required session. With this MARA prevents waste of resources (such as by avoiding sending packets to leaf nodes without active member users) and session quality degradation due to re-routing (caused by unpredictable link failures, new *join/leave* events or handovers).

For instance, whenever an ingress router detects that a session has no leaf nodes in some of the egress routers of its current multicast aggregation tree, MARA performs the following operations: i) selects another multicast aggregation tree leading to the right set of egress routers and supporting the required QoS (as deployed upon receiving the *Ri*); ii) switches the session to the selected multicast tree (by controlling the aggregation accordingly). In the case that none of the available multicast aggregation can support the QoS required by the affected session, MARA tries to readjust the current resource allocations. If the switching fails (unavailable multicast aggregation trees), the request is denied.

IV. MARA PERFORMANCER EVALUATION

The impact of MARA over the performance of the network and its comparison with MIRA is evaluated through simulations carried out using the *Network Simulator-2* (NS-2). A network topology composed by 14 nodes, interconnected by links with different capacities and propagation delay, was randomly generated by BRITE. As suggested in [9], the simulation model supports one *Expedited Forwarding* alike CoS (Premium), two *Assured Forwarding* alike CoSs (Gold and Silver), and one Best Effort class. The initialization factor is ¼ (25%), *MRth* is 20% and *CRth* is 50% of the *MRth* for each one of the 4 classes. In order to achieve all functionalities required to accomplish the evaluation, the version 2.29 of NS-2 was extended with the WFQ discipline (for QoS scheduling), PIM-SSM (for IP multicast) and MARA agents.

The experiments were repeated 10 times, and simulated a large number of multi-user UDP sessions (1,000) sent from the same ingress router. The sessions have a lifetime that varies from 20s (short live) to 120s (long live), and have a constant bit rate of 224Kb/s, to emulate a three flows scalable rate composition (32Kb/s, 64Kb/s and 128Kb/s) [10]. A Poisson distribution generates session requests and releasing requests (according to the lifetime of each session) from the beginning up to the end of the simulation (120s). MARA is configured to over-provision multicast trees with a maximum limited number of 6 hops. Before carrying out MARA evaluations, we deployed simulation studies to analyze the behavior for selecting multicast aggregation trees with different configuration for the maximum limit of hops. The results showed that multicast trees have in average between 3 and 11 hops, and on previous work from other researchers attesting that 80% of intra-domain shortest paths have 4 hops or less [11], [12].

The current simulations confirm previous results in terms of number of traversed hops, showing that the selected multicast aggregation trees comprise a maximum of 5 hops. The selected multicast trees correspond in average to 5.4% of all available trees.

In order to analyze MARA performance and impact in the network, we traced the signaling load throughout the network and the rate of *RESERVE* messages during the entire simulation. Fig. 4 shows the accumulative signaling load introduced in the set of experiments.

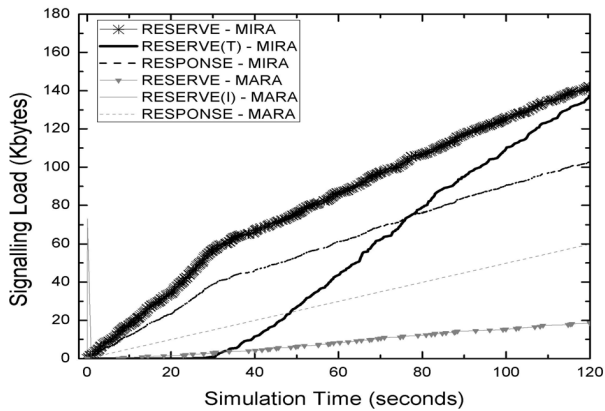

Fig. 4. Signaling load of MARA in comparison with MIRA

The traced results show an average rate of 141.96KB and 18.76KB of *RESERVE* message for MIRA and MARA, respectively. The per-flow resource allocation signaling of MIRA is the reason for its large signaling load. In contrast, MARA signaling occurs mostly during the system initialization phase (73.12KB) for *RESERVE(I)*), being the signaling load during the simulation resultant of the dynamically update of over-reservations (maximum of 14.59KB of *RESERVE*). As a consequence, MIRA also generates more *RESPONSE* messages (102.80KB) than MARA (59.85KB). Furthermore, per-flow releasing operations (triggered by *RESERVE(T)* messages, and invoked whenever a multi-user session ends) introduce an average of 137.46KB for *RESERVE(T)* MIRA messages. MARA has no *RESERVE(T)* messages, since the end of multi-user sessions grants the associated CoS with surplus resources, instead of releasing them, as occurs with MIRA. Hence, MARA minimizes in 60.33% the signaling load of MIRA.

Additionally, simulation results reveal that MARA takes approximately 0.04s to successfully initialize the system (signaling pike in the beginning of the simulation), while the signaling load generated do not exceed 8% of the link capacities throughout the network. Thus, we attested that the system initialization, performed by MARA, does not damage the network performance.

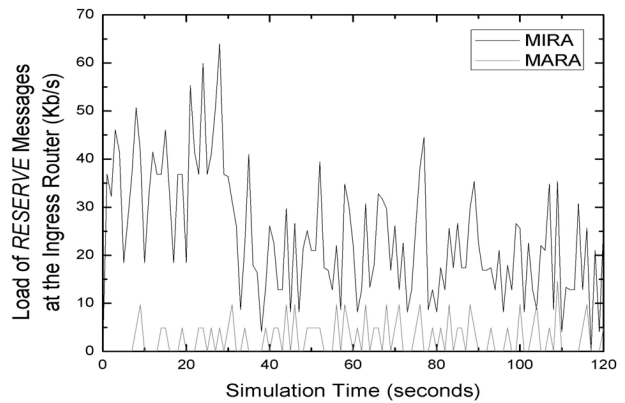

Fig. 5. Load of *RESERVE* messages

Fig. 5 shows the load of *RESERVE* messages measured in the ingress router of the simulation model. The traced results expose that within all admitted requests, MIRA established 65.9% and MARA 65.7% (less 2) multi-user sessions in the set of experiments. The session blockings in MIRA experiments are of 32.8% for *Premium*-alike, 32.5% for *Gold*-alike and 34.6% for *Silver*-alike, whereas in MARA are of 27.9%, 30.02% and 41.9% respectively. The static per-class *MRth* maintenance of MIRA lets that all CoSs experience similar session blockings (since the CoSs are also assigned with the same *MRth*). In contrast, the dynamic re-sizing of CoS's *MRth* allows MARA granting congested CoSs with more resources to establish session requests more frequently received (*Premium*- and *Gold*-alike in this case). Moreover, this dynamic re-adjustment allows MARA preventing waste of resources and improving bandwidth usage. For instance, consider a scenario where a CoS is more demanded than the others, and that its *MRth* is not enough to admit all associated requests. Thus, MARA is supposed to provide more resource allocation efficiency than MIRA.

In addition, MARA allowed the admission of 74.9% without any signaling exchange. This is attested by the number of idle signaling periods, since MARA only signals the network upon computing a new over-reservation for CoSs that are experiencing some congestion, situation that occurred for 25.1% of the admitted sessions. The simulation result shows that MIRA introduces significantly more *RESERVE* messages than MARA (23.75Kb/s and 2.83Kb/s on average respectively) during the entire simulation, letting us to conclude that MARA significantly minimizes the processing load when compared with MIRA.

In addition to the signaling and processing costs, the state stored in the system is also an important scalability measure. In what concerns state, MIRA and MARA similarly control sessions and bandwidth reservations. The major difference resides in the way multicast state is controlled. Whereas MIRA allocates multicast channels for each session-flow during the entire simulation time, MARA over-provisions multicast aggregation trees at bootstrap. In order to examine the accuracy of MIRA and MARA we analyzed the manipulation of multicast state along the simulation time. Simulation results show when and how much state is currently being handled. The multicast state in the ingress router is the focus of this analysis, since its statefull nature requires more attention due to the additional processing/state overhead in comparison with core routers. Fig. 6 illustrates the multicast state stored in the system in function of the simulation time.

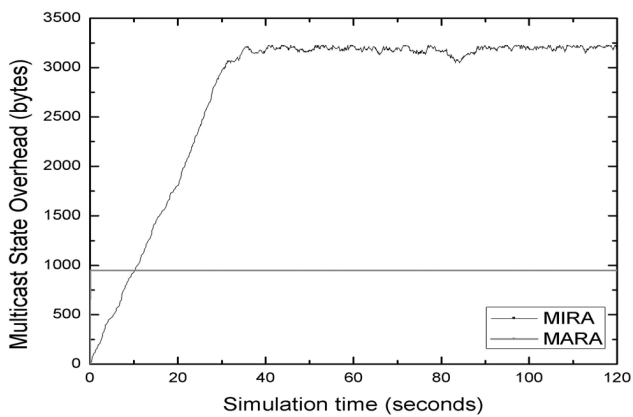

*Fig. 6. Multicast state behaviour*

The results reveal that MARA minimizes in 67.5% (231 trees) the averaging amount of multicast state in relation to MIRA. From second 32 up to the end of the simulation, the amount of multicast state handle by MIRA has no significant variation due to the alternate reservation and releasing events. In contrast, MARA only manipulates multicast state during the system initialization (up to 0.04s), which enforces architectural benefits in relation to per-flow multicast approaches. For instance, MARA prevents the lack of available multicast IP addresses (mainly in IPv4 environments), as well as address collisions. Those problems are influenced by unpredictable reception of *join* and *leave* requests in large-scale networks. In addition, multicast aggregation without per-flow signaling allows a significant reduction of state and signaling overhead to control multicast trees.

## V. CONCLUSION AND FUTURE WORK

This paper introduces the *Multi-user Aggregated Resource Allocation* mechanism (MARA) to coordinate the over-provisioning of bandwidth and multicast trees to be used by multimedia sessions shared by more than one receiver (multi-user sessions). The proposed mechanism aims to mitigate the scalability problems of per-flow signaling approaches, such as the *MultI-service Resource Allocation* mechanism (MIRA) previously proposed by the authors. The combination of over-provision of resources (bandwidth of DiffServ classes and multicast trees) with admission control allows MARA to setup a significant number of sessions (74.9%) without any signaling in the network. Moreover, the performance limitations of IP multicast (state and maintenance signaling) are overcome by using multicast aggregation. Finally, waste of resources due to over-reservations is prevented by dynamically re-adjusting resources and controlling session connectivity on-demand.

Simulation results prove the benefits of MARA in relation to MIRA. Although MARA performs all over-reservations in-advance, after the system bootstrap, simulation results show that system initialization takes only 0.04s and consumes no more than 8% of the link capacity. During the simulation MARA reduces in 60.33% the signaling load in relation to MIRA. In what concerns state overhead, MARA reduces the state of the ingress router (the most problematic point) in 67.5% in comparison to MIRA.

REFERENCES

[1] A. Neto, E. Cerqueira, A. Rissato, E. Monteiro, and P. Mendes, "A Resource Reservation Protocol Supporting QoS-aware Multicast Trees for Next Generation Networks", In *Proc. 12$^{th}$ IEEE Symp. on Computers and Communications*, Aveiro, Portugal, Jul. 2007.

[2] P. Pan, E. Hahne, and H. Schulzrinne, "BGRP: A Tree-Based Aggregation Protocol for Inter-domain Reservations", *Trans. of Communications and Networks Journal*, vol. 2, pp. 157-167, Jun. 2000.

[3] R. Sofia,r R. Guerin, and P. Veiga, "SICAP, a Shared-segment Inter-domain Control Aggregation Protocol", in *Proc. Int. Conf. in High Performance Switching and Routing*, Turin, Italy, Jun. 2003.

[4] R. Bless, "Dynamic Aggregation of Reservations for Internet Services", in *Proc. 10th Conf. on Telecommunication. Systems –Modeling and Analysis*, vol.1, pp.26-38, Monterey - California, USA, Oct. 2002.

[5] P. Pinto, A. Santos, P. Amaral, and L. Bernardo, "SIDSP: Simple Inter-domain QoS Signaling Protocol", in *Proc. IEEE Military Communications Conference*, Orlando, Florida, USA, Oct. 2007.

[6] T. Braun, V. Arya, and T. Turletti, "A Backup Tree Algorithm for Multicast Overlay Networks", in *Proc. Networking*, Waterloo, Canada, May 2005.

[7] M. Kodialem and T. Lakshman, "*Dynamic routing of bandwidth guaranteed multicasts with failure backup*", in *Proc. IEEE INFOCOM*, New York, USA, Jun. 2002.

[8] S. Hanna, B. Patel, and M. Shah, "Multicast Address Dynamic Client Allocation Protocol (MADCAP)", *IETF RFC* 2730, Dec. 1999.

[9] Z. Di and H. Mouftah, "Performance Evaluation of Per-Hop Forwarding Behaviours in the DiffServ Internet", in *Proc. IEEE Symposium on Computers and Communications*, Antibes-Juan les Pins, France, Jul. 2001.

[10] K. Rose and S. Regunathan, "Toward optimality in scalable predictive coding", *IEEE Transaction on Image Processing*, Vol. 7, pp. 965-976, Jul. 2001.

[11] T. Wong and R. Katz, "An analysis of multicast forwarding state scalability", *In Proc. 8$^{th}$ Int. Conf. on Network Protocols*, Osaka, Japan, Nov. 2000

[12] A. Shaikh, R. Tewari, and M. Agrawal, "On the Effectiveness of DNS-based Server Selection", *IBM Research Report*, 2000.